\RequirePackage{lineno}
\documentclass[twocolumn,letterpaper,aps,prl,superscriptaddress,showpacs,floatfix]{revtex4}

\usepackage{graphicx}	

\usepackage{xspace}	
\usepackage{amsmath}

\begin{document}


\title{Understanding the Effect of Atmospheric Density on the Cosmic Ray Flux Variations at the Earth Surface}

%

\author{Mathes Dayananda, Xiaohang Zhang, Carola Butler and Xiaochun He}

\date{\today}

\begin{abstract}

We report in this letter for the first time the numerical simulations of muon and neutron flux variations at the surface of the earth with varying air densities in the troposphere and stratosphere. The simulated neutron and muon flux variations are in very good agreement with the measured neutron flux variation in Oulu and the muon flux variation in Atlanta.  We conclude from this study that the stratosphere air density variation dominates the effects on the muon flux changes while the density variation in troposphere mainly influences the neutron count variation.  These results pave a new path for systematically studying the global temperature evolution using worldwide cosmic ray data.
\end{abstract}

	



\maketitle

Sporadic weather patterns occur so frequent in recent years. According to NASA,  the year 2012 was the ninth warmest year since 1880.
The temperature at the surface of the earth has increased by ${\sim}0.6^{\circ}$C during the last century \cite{stozhkov}. 
Currently, there are two competing interpretations of the causes for the global warming:  natural forcing such as solar influences on the climate, and influences of anthropogenic activities. The studies done by Svensmark \cite{svensmark} and others \cite{shaviv2005,ollila}, on the other hand, indicate that the galactic cosmic rays (GCR) may play a significant role on the temperature variation of the earth.      

Over the past decades, quite a few studies reported on the correlations between the earth weather and cosmic ray flux \cite{svensmark,MarshSvensmark,CLOUD,Kirkby,Lu}. 
 It is showed that an 11-year average of northern hemispheric land and marine temperature is more closely correlated with the measured variation in cosmic ray flux \cite{svensmark}. Furthermore, there exists a causal relationship between GCR and low cloud coverage ($< 3.2 $ km) which suggests that the formation of the low clouds is influenced by the atmospheric ionization produced by GCR \cite{MarshSvensmark}. Clouds play a significant role on the earth's radiation budget by reflecting the incoming short waves and trapping the outgoing long waves. This causal relationship reflects that the variation in GCR may indirectly influence the earth's temperature. 

\begin{figure}
\centering
\includegraphics[width=0.93\linewidth]{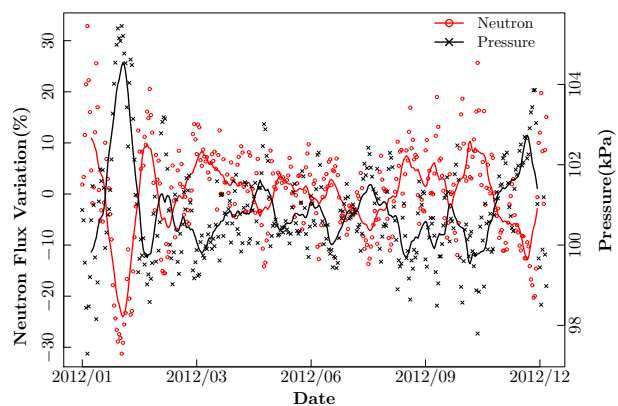}
\caption{\label{fig:oulu} (color online)  Neutron daily counts percentage variation (red) and atmospheric pressure (black) measured at the Sodankyla Geophysical Observatory in Oulu, Finland in 2012 (http://cosmicrays.oulu.fi/).  The solid lines are two-week moving averages. 
}
\end{figure}

\begin{figure}
\centering
\includegraphics[width=0.93\linewidth]{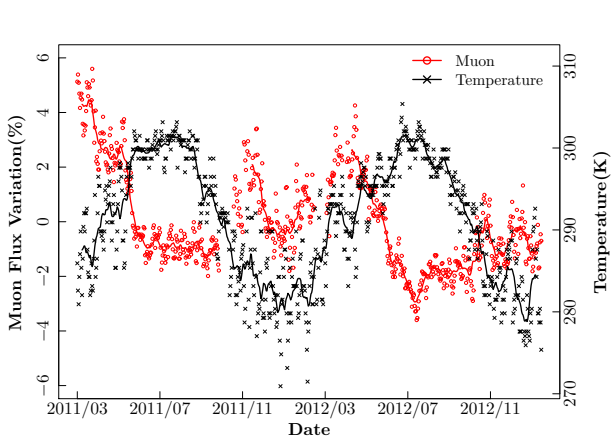}
\caption{\label{fig:pot} (color online)  Muon daily counts percentage variation (red) and temperature (black) measured in Atlanta, Georgia from 2011 to 2013.  Temperature data is from Atlanta Fulton weather station (http://www.wunderground.com/). The muon counts were recorded with a liquid scintillator detector installed on campus at Georgia State University. The solid lines are two-week moving averages. 
}
\end{figure}

While the true impact of cosmic rays on the earth climate change is currently far from conclusive, continued efforts of long-term monitoring of cosmic ray flux variations are imperative. This study also requires a quantitative understanding of the influence of atmosphere air density fluctuations to the cosmic ray flux.   Figure~\ref{fig:oulu} shows the neutron daily counts percentage variation (blue) and pressure (red) measured at the Sodankyla Geophysical Observatory in Oulu, Finland in 2012.  The Sodankyla Geophysical Observatory has been recording the neutron counts since 1964.
At Georgia State University, a seasonal muon flux variation is observed  as shown in Fig.~\ref{fig:pot}. The data were recorded from March of 2011 to January of 2013. Also shown in Fig.~\ref{fig:pot} is the ground temperature recorded during the same period. The data shows a significant drop of muon counts during the summer time while a higher counting rate seen in winter similar to the trend observed by the IceCube Neutrino Observatory \cite{iceCube} and others \cite{Bertaina}.

It is therefore important to precisely understand how both the muon and neutron flux is attenuated in different layers of the atmosphere. Numerical simulations of muon and neutron flux variations at the surface of the earth have been carried out with varying air densities in the troposphere and stratosphere. The simulation software is developed based on the Geant4 package \cite{G4}. Results are reported in this letter.

In the simulation setup, a column of air $100$ km in height and $50$ km in diameter (composed of 70\% Nitrogen and 30\% Oxygen) is configured.
The atmospheric air density is parameterized in the following equation \cite{nasaDenMod}.
\begin{equation} \label{eq:rho}
  \rho = \frac{P}{0.2869 (T+273.1)} 
\end{equation}
where $\rho$ is the air density in kg/m$^3$, $P$ is the pressure in kPa, and $T$ is the temperature in celsius.  





The primary cosmic particles impinging on the top of the earth's atmosphere consist of $79$\% protons and a small fraction of alpha particles and heavier nuclei \cite{pdg_cosmic}. In the present work, only primary protons are included in the simulation. The protons are launched vertically downward at the top of the air column with an energy distribution as described in \cite{pdg_cosmic}. 

Based on the data provided by NASA \cite{msis}, the air density in the stratosphere is larger in summer time and smaller in winter, which is opposite to the air density variation in the troposphere. This air density variation is modeled in our simulation by scaling the air density to match the seasonal variation in order to study its  effect on neutron as well as muon flux changes measured at the surface of the earth.

\begin{figure}
\centering
\includegraphics[width=0.95\linewidth]{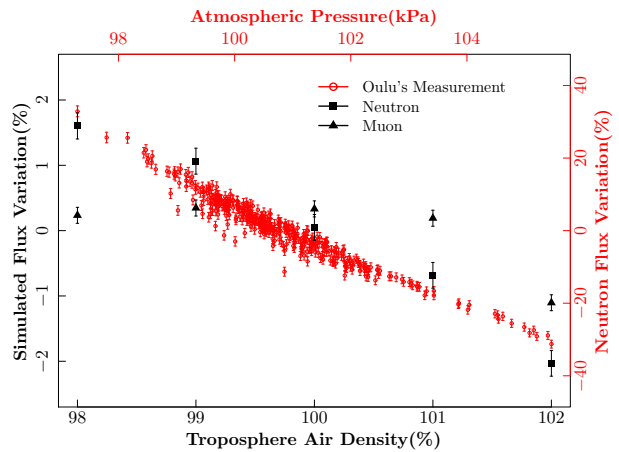}
\caption{\label{fig:neutron} (color online) Simulated percentage variation of ground level neutron and muon counts as function of the percentage variation of the troposphere density (black).  Also shown in this figure is the daily neutron flux percentage variation as a function of the atmospheric pressure from Oulu (red).}
\end{figure}

The empirical atmospheric data \cite{msis} shows a $\pm2.5\%$ maximum variation of the troposphere density between winter and summer.  This variation in density is modeled in our simulation by scaling the troposphere density from 98\% to 102\% relative to the yearly average while keeping the same stratosphere density. Figure~\ref{fig:neutron} shows the simulated neutron counts (black solid squares) variation at the ground level as a function of the percentage variation of the troposphere density.  
Also shown in Fig.~\ref{fig:neutron} is the measured neutron counts variation (red circles) from Oulu observation in 2012.  The simulated result shows a very consistent trend with the data.  This indicates that the neutron flux variation is primarily influenced by the modulation of the air density in the troposphere.   One should not expect a perfect correlation between the measured data and the simulated results for two reasons: (1) the simulation does not include the Oulu detector acceptance and efficiency; (2) a constant primary cosmic ray flux is used in the simulation.  

While there is a strong anti-correlation between the troposphere air density and the neutron flux,  there is little effect on the muon flux from the troposphere air density variations as shown in Fig.~\ref{fig:neutron} in black triangle points (except a noticeable effect on muon flux changes at the higher troposphere air density range).  This is consistent with the results reported in \cite{Chilingarian} and led us to believe that the stratospheric air density may play a significant role for modulating the muon flux variation as seen in Fig.~\ref{fig:pot}.
 
\begin{figure}
\centering
\includegraphics[width=0.95\linewidth]{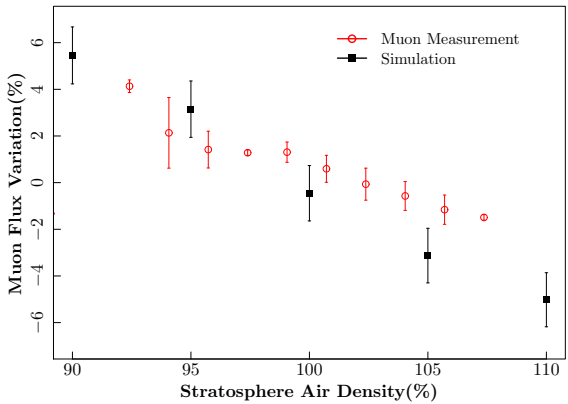}
\caption{\label{fig:muon} (color online) Ground level muon flux percentage variation as a function of percentage variation of the stratospheric air density.  The red-circle data points are from GSU measurement and the black-square from the simulated results. 
We only sample muon particles with kinetic energy $\ge 1$ GeV at ground level which corresponds to the detector threshold. 
Note that the error bars in the figure are statistical only. } 
\end{figure}

In order to make a realistic simulation study, we modeled the stratosphere density variation according to the data from the Peachtree City Observation Station (Department of Atmospheric Science, University of Wyoming).  In our simulation, stratosphere density is varied from 90\% to 110\% relative to its average while the troposphere density is kept constant. The results are shown in Fig.~\ref{fig:muon}. A significant reduction of muon flux is clearly seen with increasing stratospheric air density, which is very consistent with our measurements at GSU. Based on the results in Fig.~\ref{fig:neutron} and Fig.~\ref{fig:muon}, it is clear that the effect of density fluctuation in the stratosphere region is dominant on the muon flux variation comparing the modulation effect from the troposphere region. 

\begin{figure}
\centering
\includegraphics[width=0.98\linewidth]{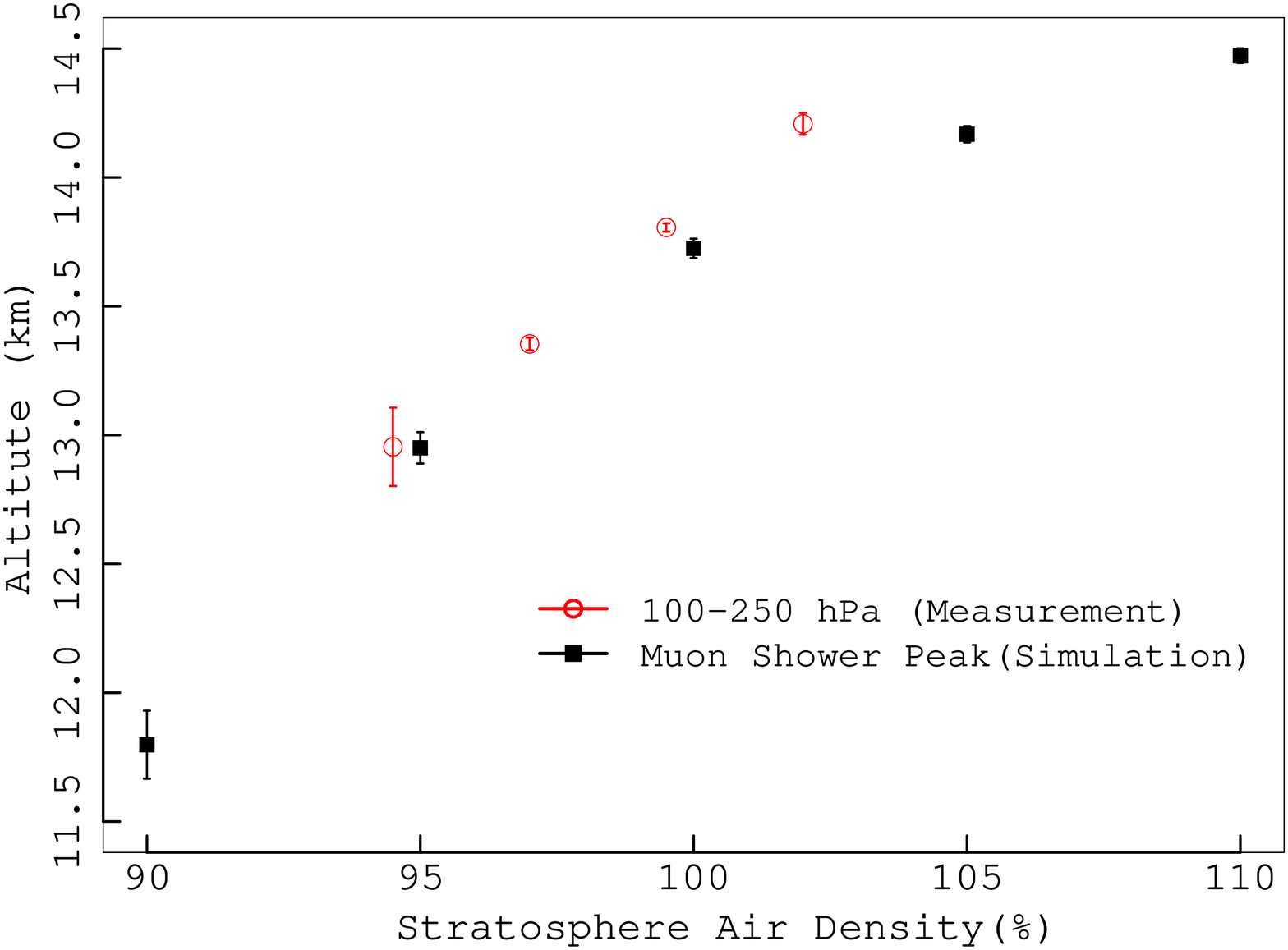} 
\caption{\label{fig:shower} (color online) Simulated cosmic ray shower maximum altitude (black square) as a function of the percentage variation of the stratosphere air density. The red circle data points represent the mean altitude of the pressure range of $100$ to $250$ hPa, which are taken from the Peachtree City Observation Station (2011 - 2013).}
\end{figure}

The modulation of muon flux from the variable stratospheric air density can be further understood by studying the cosmic ray shower maxima altitude distributions. 
Primary cosmic ray showers mainly occur where the atmospheric pressure is between $100$ and $250$ hPa, implying that there is a certain density threshold ($\sim$$0.2$ g/$m^3$) necessary for maximum likelihood of primary cosmic ray interactions. 
Figure~\ref{fig:shower} shows the altitudes of the simulated shower maxima (black squares) as a function of the stratospheric densities. Also shown in Fig.~\ref{fig:shower} is the variation of mean altitude (red circles) of the stratospheric pressure range of $100$ to $250$ hPa extrapolated from the Peachtree City Observation Station data from 2011 to 2013. Note that the larger error bars at low densities are due to low statistics. 
A remarkably good agreement is seen between the simulated shower maximum altitude distribution and the extrapolated pressure threshold region. 

It is important to point out that the stratospheric air density variation indirectly reflects the seasonal temperature variation in the stratospheric region. The results shown in Fig.~\ref{fig:muon} suggest that muon flux is primarily influenced by the seasonal temperature variation in the stratosphere. 


It is known that the temperature in the troposphere can fluctuate considerably within a day while the stratospheric temperature only varies seasonally unless there is a sudden stratospheric warming \cite{minos}. The primary cosmic ray particles mainly interact with the stratospheric nuclei and generate secondary cosmic ray particles at an altitude between $12$ and $15$ km. The mesons produced in these cosmic showers can either interact with the atmosphere or decay into muons. During the summer time  due to the expanding atmosphere cosmic ray showers occur at higher altitudes. This means that the muons travel further to reach the surface of the earth and are more likely to decay leading to a lower muon rate in summer and a higher rate during winter.   

The simulated results in Fig.~\ref{fig:muon} are obtained by only varying the stratospheric density and keeping a constant tropospheric density. However, empirical atmospheric data \cite{msis}  show that the stratospheric and tropospheric densities vary inversely during summer and winter. The measured muon rate is higher compared to the simulated muon rate in summer times as a result of less modulation by the low troposphere density. This trend is also consistent with our muon simulation results shown in Fig.~\ref{fig:neutron}.


In summary, we report in this letter for the first time the simulation studies of the effects of the tropospheric and stratospheric density variations  on the cosmic ray muon and neutron flux  reaching to the surface of the earth. Our simulation results show that the density variations in the troposphere mainly influence the neutron flux while its influence on the muon flux is relatively insignificant.  On the other hand, the stratospheric density dominates the muon flux which is in remarkably good agreement with observed seasonal variation of muon flux.   It is, therefore, very important to have a longterm monitoring of both muon and neutron flux simultaneously at a global scale in order to study the dynamical change of the atmosphere.

We would like to acknowledge the wealth data sets which are available online provided by the Sodankyla Geophysical Observatory (Oulu, Finland), the Peachtree City Observation Station and Weather Underground, Inc.



\begin{acknowledgments}


\end{acknowledgments}




\bibliography{fluxSimu_arXiv}   

%


\end{document}